# Bike3S: A Tool for Bike Sharing Systems Simulation

Alberto Fernández*, Holger Billhardt, Sacha Ossowski, Óscar Sánchez

CETINIA, University Rey Juan Carlos, Madrid, Spain

Tulipán s/n, 28933, Móstoles, Madrid, Spain {alberto.fernandez, holger.billhardt, sascha.ossowski, oscar.sanchezsa}@urjc.es

ORCID Alberto Fernández: 0000-0002-8962-6856,

ORCID Holger Billhardt: 0000-0001-8298-4178

ORCID Sacha Ossowski: 0000-0003-2483-9508

# Bike3S: A Tool for Bike Sharing Systems Simulation

Vehicle sharing systems, such as bike, car or motorcycle sharing systems, are becoming increasingly popular in big cities as they provide a cheap and green means of mobility. The effectiveness and efficiency, and thus, the quality of service of such systems depends, among other factors, on different strategic and operational management decisions and policies, like the dimension of the fleet or the distribution of vehicles. In particular, the problem of agglutination of available vehicles in certain areas whereas no vehicles are available in other areas is a common problem that needs to be tackled by the operators of such sharing systems. Recently, research works have been focusing on adaptive strategies to reduce such imbalances, mainly through dynamic pricing policies. However, there is no best operational management strategy for all types of bike sharing systems, so it is of foremost importance to be able to anticipate and evaluate the potential effects of such operational management strategies before they can be successfully deployed in the wild. In this paper we present Bike3S, a simulator for a station-based bike sharing system. The simulator performs semi-realistic simulations of the operation of a bike sharing system in a given area of interest and allows for evaluating and testing different management decisions and strategies. In particular, the simulator has been designed to test different station capacities, station distributions, and balancing strategies. The simulator carries out microscopic agent-based simulations, where users of different types can be defined that act according to their individual goals and objectives which influences the overall dynamics of the whole system.

## 1    Introduction

Nowadays, in urban mobility there is a trend towards limiting the use of vehicles with combustion engine, especially if they are used for private transportation. The aim is to reduce their environmental impact as well as the occupation of public spaces. At the same time, citizens are demanding flexible, individualised mobility solutions that adapt to their specific needs at any moment, and that are aligned with their environmental, health and cost concerns. As a result, there is a growing deployment of sustainable mobility systems, with zero or low emissions and shared vehicles.

In this context, many big cities around the world are encouraging cycling mobility among their citizens, not only by improving the cycling infrastructure (bicycle lanes etc.), but also by installing bike-sharing systems (BSS). Some BSS (e.g. *BikeShare* in Seattle or DB *Callabike* in Munich) allow citizens to pick up and return a bike at any location within a certain area in town, but most systems are station-based, i.e. they rely on a set of rental stations with fixed locations. Several BSS are quite sizeable, reaching about 20000 rental bicycles in Paris (Vélib) and more than 50000 in Hangzhou, China (Shaheen, Guzman, & Zhang, 2010). Studies have shown that the adoption of new BSS depends strongly on the perception of their efficiency and positive effects, individually and within the community.

Both the quality of service offered to the citizens, as well as the (economic and environmental) cost of running a BSS depends strongly on taking proper management decisions. This does not only include *strategic* choices regarding the positioning and dimensioning of rental stations, the selection of adequate bicycle models, etc., but also on *operational* decisions. Resources in a BSS are limited, and not being able to find an available bike at some stations, or not being able to return it to another one due to the lack of free slots, are events that can severely deteriorate user experience in a BSS. ICT-based solutions that allow users to *reserve* bicycles or free slots at some stations palliate this problem, but proper bike balancing mechanisms are needed to attack the problem at its core.

Traditionally, bike balancing has been done by trucks that transport bicycles from some stations to others. Some research has focused on optimizing static balancing problem (Chemla, Meunier, Pradeau, Calvo & Yahiaoui, 2013; Forma, Raviv & Tzur, 2015), where the routes of trucks at night or off-peak periods are optimised. More recently, the dynamic version of the problem has been considered, which involves predicting the demand on each station in the next period and optimizing the distribution of bikes in stations so as to maximise the number of trips (i.e. reduce the number of "no-service" situations) (Contardo, Morency, & Rousseau, 2012; O'Mahony, Shmoys, 2015; Schuijbroek, Hampshire, & Van Hoeve, 2017). While those approaches only consider trucks as the means to rebalance the bike-sharing system, there are other approaches that try to incentivise bike users to collaborate in the system rebalancing (Chemla, Meunier, Pradeau, Calvo & Yahiaoui, 2013; Fricker & Gast, 2012; Pfrommer, Warrington, Schildbach & Morari, 2014; Waserhole & Jost, 2016). For this purpose, prices are commonly used as incentives. For example, in the city of Madrid BiciMAD[1] grants discounts over the usual rental price if a user picks up a bike at a station that is almost full with parked bicycles or if she returns it to a station with many empty slots. Experiments with approaches that modify rental prices dynamically, based on the BSS load situation, and with social stimuli to achieve voluntary travel behaviour change of BSS clients have also been reported.

Still, it is known that there is no best management strategy for all types of BSS, so it is of foremost importance to be able to anticipate and evaluate the potential effects of such operational management strategies in a particular BSS before they can be successfully deployed in the wild. For this purpose, firstly, a particular BSS needs to be modelled at sufficient level of detail, including the positions and size of rental stations, a town cycling and street network, different user demand patterns, etc. And, secondly, the action space of BSS users and their behavioural choices need to be realistically modelled, in particular with regard to economic and social stimuli. To the best of our knowledge, there are currently no BSS simulators that fully account for these requisites. In this paper we present Bike3S, a simulator for a station-based bike sharing system. The simulator performs semi-realistic simulations of the operation of a bike sharing system in a given area of interest and allows for evaluating and testing different management decisions and strategies. In particular, the simulator has been designed to test different stations capacities, station distributions, and balancing strategies. The simulator carries out microscopic agent-based simulations, where users of different types can be defined that act according to their individual goals and objectives which influences the overall dynamics of the whole system.

---

[1]Public bike sharing system of Madrid (Spain): https://www.bicimad.com/

The rest of the paper is organised as follows. In Section 2 we discuss related work on bicycle sharing and simulation in general. Sections 3 focuses on station-based BSS that we are particularly concerned with. Section 4 discusses the general architecture of the Bike3S simulator and its different elements in detail. In Section 6 we put forward different use cases, including the application of Bike3S to BSS scenarios for the cities of Madrid and London. Section 7 summarises our work, outlines the lessons we have learnt, and points to future lines of work.

## 2    Related works

Microscopic simulation has been successfully used in the transportation community for many years. Specially, several well-known tools were created for traffic simulation, such as SUMO (Krajzewicz, D., Erdmann, J., Behrisch, M., & Bieker, 2012), MATSim (Horni, Nagel & Axhausen, 2016) or PTV Vissim (Fellendorf, 1994).

The state of the art on simulating bike sharing systems is not as advanced as those works on traffic simulation mentioned above. Nevertheless, the increase of popularity of this kind of transportation means is provoking that the research community is paying attention to this area.

Chemla et al. (2013) presented a discrete-events open-source simulator (OADLIBSim) for evaluating their proposed algorithms for dynamic balancing. The user behaviour for choosing stations combines walking and riding distances. If there are no available bikes at a station the user may try in a different one up to a limited number of times. The number of satisfied users is taken as a quality measure to compare different balancing strategies. Unfortunately, the simulator is no longer available. Caggiani, and Ottomanelli (2012) proposed a model and simulator with the goal of optimising bikes repositioning and routes of carrier vehicles. In their model, a day is divided into discrete time intervals. At each interval the O-D demand is considered, and new states of stations are calculated. A user that fails to hire a bike (empty station) is removed, and if the destination is full, she has to wait to return her bike.

Romero, Moura, Ibeas and Alonso (2015) combined the use of cars, buses and bikes into an integrated transportation model. Their focus is on modelling users' transportation choices based on urban transportation infrastructures so as to analyse their effect on urban mobility. Our objective is more focused on bike sharing infrastructures, mainly oriented towards balancing strategies.

Soriguera, Casado and Jiménez (2018) developed a very complete simulation model and tool with the goal of supporting the decisions regarding deploying a BSS. The tool is based on Matlab. They include repositioning trucks and the possibility of using electric bikes. Users take and returns bikes as near as possible to their locations (it is assumed that users know availability). This work is probably the closest to ours. We keep tracks and electric bikes as part of our future lines. Our simulator is more general leaving most of the users' decisions (e.g. next station) up to the specific user implementation, as we will see later.

In general, we present in this paper a BSS simulator that is highly configurable with many parameters at different levels: global, user models, user generation, balancing strategies. User types and strategies can be easily extended by developers. In addition, it includes user interfaces for configuring, simulating and visualisation. To the best of our knowledge, there are not any other BSS simulators as complete as the one presented in this paper.

## 3    Station-based Bike Sharing Systems

In this section we describe the general functioning of a station-based bike sharing system. The description is inspired by the functioning of the BICIMAD system in the city of Madrid in Spain. However, it is very general and fits many similar systems in other cities.

A station-based bike sharing system consists of a set of docking stations distributed in an area of interest (typically a city) and a set of bikes that can be taken from or returned to a station. Stations are at fixed positions and have a set of slots where bikes can be plugged into. The number of operative slots of a station represents its capacity. At a given moment, a station may have broken slots, or some slots may be occupied by broken bikes. We do not deal with such cases. However, they could be easily modelled as a decrease in the station's capacity.

In general, when a user wants to use the system she would go to a station, take a bike and return it after some time at another station. A user may go directly to a station close to her position or she could consult some kind of information or recommendation service to find the closest station. Furthermore, we assume that there is some kind of registry of users (e.g., long-term contracts or occasional users) and also that there is some type of payment involved in the use of the system. However, these aspects are not included in our simulation tool.

Besides the possibility of simply taking an available bike at a station, we consider that it is possible to reserve a bike (or a slot for returning a bike) at a station. Usually, systems that allow for reservations block the reserved bike (or slot) for a fixed amount of time, such that no other user can use it. In the BICIMAD system in Madrid, this reservation time is 20 minutes. Thus, any bike in the system would be either "in use", "available at a station" or "reserved at a station". Regarding the empty slots, they may be either "available for use" or "reserved".

Depending on the situation of the infrastructure (e.g., the availability of bikes or slots), as well as on the users individual objectives, each user that enters the bike sharing system follows a path from an initial state to an end state within the diagram presented in Fig. 1.

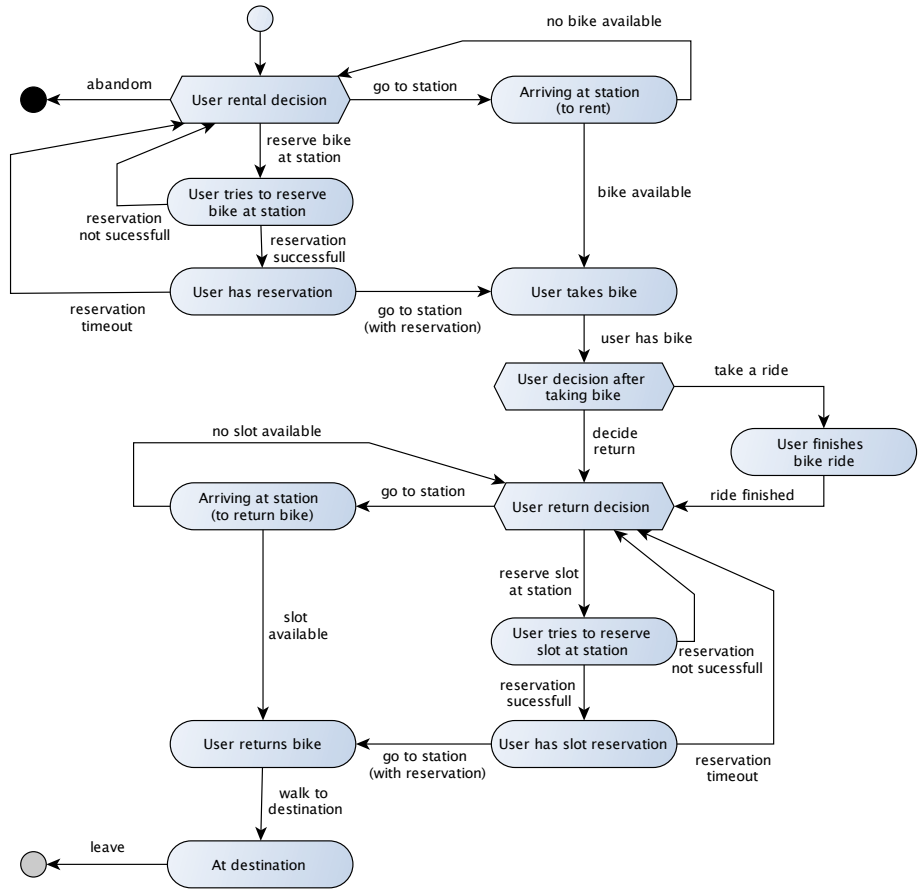

Fig. 1. User life cycle state diagram.

A user appears in the system at some time and position with the intention to rent a bike. In general, users appear at any locations in the city, not necessarily at stations. After appearing, the user handles a rental decision with the following possible outcomes: i) reserve a bike at some station, ii) go directly to some station in order to rent a bike, or iii) abandon the system. For reserving and/or finding stations, the user may use some available information or recommendation system. If the user decided to reserve a bike, she tries to do so (possibly using some application). If the reservation has been successful, the user goes to the station where she reserved a bike. If the reservation was not successful, the user again has to decide what to do. An active reservation may finish after some timeout period, without the user having reached the station. We assume that the user is informed about such a timeout (e.g., via an app) and thus, has to retake again her rental decision. In case a user arrives at a station (without a reservation), it may happen that there are no bikes available when the user arrives. Also, in this case, the user has to decide again what is her next step. As it can be seen in the figure, the user has always the choice to abandon the system without taking a bike. Usually, a user would abandon if she does not find a bike after trying in one or more stations or if there is no bike available at a station closed to her current position.

Once the user is able to take a bike (either with or without a reservation), she will decide whether to take some ride in the city or to go directly towards a destination point. In the second case, the user will directly take the "return decision" whereas in the first case she will decide on returning the bike after the ride has finished. In the "user return decision", she will select a station to leave the bike (usually near her destination point) and decide to either reserve a slot at that station or to go there without a reservation.

The reservation of slots is treated similarly as the reservation of bikes. The user re-decides if a reservation intent has not been successful and also if the reservation timeout has occurred. If the user goes directly to a station (without a reservation), it could be possible that, after arriving, there is no empty slot available. Also, in this case the user has to reconsider its return decision. Eventually, the user reaches a station with an available slot (with or without reservation) and can return the bike, walk to her final destination and leaves the system. It should be noted that a user can not abandon the system when he has a bike. He first has to return that bike. That means, we do not consider possible malicious behaviours of users.

It should be noted that the "user rental decision" and the "user return decision" may be repeated during the live cycle of the user in the system. Obviously, a user knows about past events (e.g., failed reservations, failed bike rentals, etc.) and, thus, may decide differently in these decision processes in different situations.

## 4 Architecture

In this section we describe the architecture of the Bike3S simulator, whose main building blocks are depicted in Fig. 2.

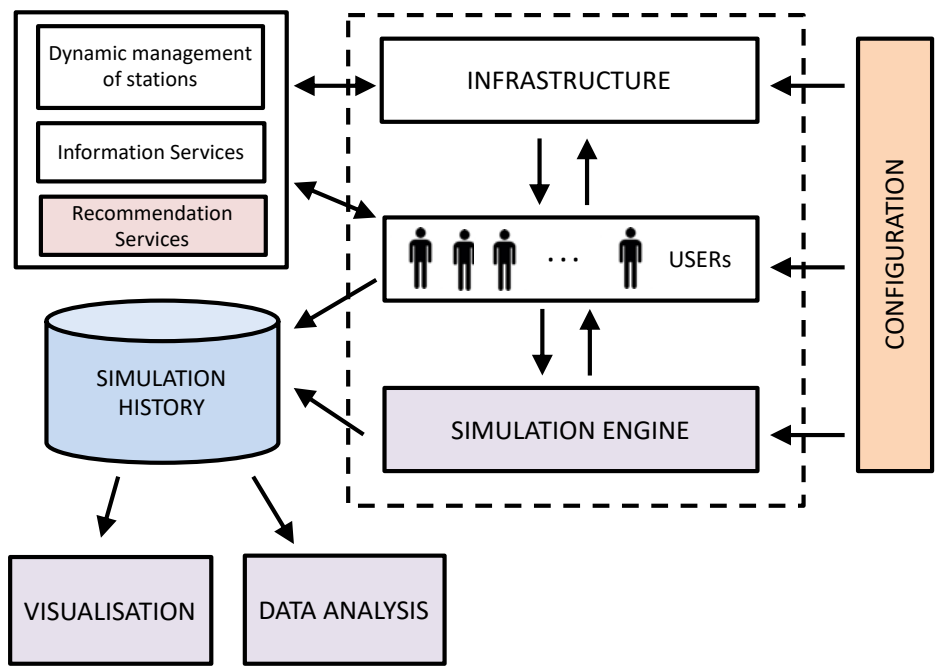

Fig. 2. Simulator architecture.

The core of the simulator consists of the simulation engine that manages the users as well as the bike sharing system infrastructure.

The bike sharing *infrastructure* represents the physical entities of the bike sharing system, i.e. stations and bikes, in the simulator. It contains information about the location of stations, number of available bikes and slots, etc. and of the bikes and slots (their current states). In the current version of the simulator, we do not distinguish among specific bikes and/or slots. However, it would be easy to extend the simulator in the future to account for characteristic like bike usage statistics, maintenance, load level (if electric), etc. The infrastructure is in charge of managing bike rentals and returns as well as bike and slot reservations.

A simulation usually involves several *users*, possibly of different types. During the simulation, users interact with the infrastructure to take/return bikes or make reservations of bikes or slots. In order to take their decisions, users may access external information or recommendation systems. In the figure we represent three kinds of external services: (i) *dynamic management of stations* refers to operator actions that would modify something in the infrastructure, including fares, for example; (ii) *information services* provide (objective) information to users about the current state of the fleet, such as the number of bikes in a station, closest station with available bikes, distance to a station, etc.; and (iii) *recommendation services* provide more processed information, which may include suggestions or requests to go to specific stations because it is better to keep the bike fleet balanced.

The initial simulation setup is accomplished through a *Configuration* module that is in charge of specifying global simulation parameters, the initial situation of stations, and the generation of users. During a simulation, the simulation engine stores the necessary information for all the events that have occurred in external files (*Simulation History*). These files can be used later to visualise the simulation in a graphic interface (*Visualization*) or to analyse the performance of simulated system with the *Data Analysis* module.

The general architecture decouples the simulation itself from the *Configuration*, *Visualization* and *Data Analysis* modules. The connection between these modules is accomplished with files in the Json format. This allows for an easy creation of other modules that could be integrated into the architecture.

The simulation core (backend) has been implemented in Java. For the user inter-faces (frontend) we used Electron and Typescript, while Angular was used for the visu-alization component.

## 4.1    *Simulation Engine*

There are two main approaches to implement agent-based simulators: i) continuous sim-ulation, and ii) event-based simulation. In continuous simulation, a fixed time-step is defined and the system executes the agents' actions (if any) and updates the state of the environment accordingly at each time step. In many steps there might not be no action nor any changes with respect to the previous state. On the other hand, in event-based simulation, the state of the environment is only updated at the precise time some event occurs. The different simulated entities may fire new events, which are processed by the simulator in sequential order.

In our case, we follow an event-based approach, where the events correspond to the different states of the live cycle of users in the system (as shown in Fig. 1).

Table 1 shows a brief explanation of each event type. Although not detailed there, each event type has different parameters: at least the time instant, and the user in-volved. Some events have additional parameters (involved station, reservation, …), de-pending on the event type.

Table 1.Event types.

| Event type | Description |
|---|---|
| User Appears | Represents the user appearance in the system. |
| User Decides Rental | The user takes the "user rental decision", deciding to abandon, go directly to a station or to try a reservation. |
| User Arrives at Station (to rent) | The user arrives at a station to rent a bike. It is not sure that the user will be able to take a bike: it depends on if the station has bikes available |
| User Tries to Reserve Bike | A user tries to make a reservation of a bike at a station. The reservation intent may be successful or unsuccessful. If unsuccessful, the user again takes the "user rental decision". If the reservation was successful, the user has a reservation. |
| User Has Bike Reservation | If a user has a bike reservation, he walks towards the corresponding station. The reservation may finish with a timeout before the user arrives at the destination station. |
| Bike Reservation Timeout | A bike reservation has expired before the user arrived at the station. The user will have to take again the "user rental decision". |
| User Takes Bike | The user takes a bike from the station and performs the "user decision after taking bike". There are two options: either to take a ride or to decide to go to some station to return the bike. |
| User Finishes Ride | The user has cycled somewhere and now has decided that he wants to go to a station to return the bike. |
| User Decides Return | This event corresponds to the "user return decision". There are two options: try to reserve a slot at the desired return station or to go directly to the desired return station. |
| User Arrives at Station (to return bike) | The user arrives at a station to return the bike. As the station may have no available slots, it is not sure the user will be able to return his bike there. In this case, he would have to take again the "user return decision". |
| User Tries to Reserve Slot | A user tries to reserve a slot at a station. The reservation intent may be successful or unsuccessful. If unsuccessful, the user again takes the "user return decision". If the reservation was successful, the user has a slot reservation. |
| User Has Slot Reservation | If a user has a slot reservation, he cycles towards the corresponding station. The reservation may finish with a timeout before the user arrives at the destination station. |
| Slot Reservation Timeout | The slot reservation has finished before the user arrives at the station. |
| User Returns Bike | The user returns the bike at the station and walks to his final destination. |
| User Arrives at Destination | The user has reached his destination and leaves the system. |
| User Leaves System | The user leaves the system. |

The simulation engine operation is based on a priority queue of events (ordered by the time instant they happen). Initially, for each user to be simulated an event of type "*User Appears"* is created with the time and location indicated in the users' configuration file. Then, the events are processed in order. The execution of events may require some decisions to be made by the user, and usually creates and inserts new events into the queue.

Movements of users in the system are simulated using an external routes server[2]. However, the simulator does not update continuously the position of the users. Instead, a new event would be created with the time at which the user would arrive at its destination. Fig. 3 shows an example, where there are several events that must occur at times 100, 120, 340, 710, 800 and so on. We simplified the information shown in the queue. The next event to be processed is *EV(UserAppears, user1,t=100, pos1)* states that *user1* appears at instant t=100 in *pos1*. When this event is processed, *user1* takes the "user rental decision". Let´s suppose she decides to go directly to station 55. After the decision, the route and arrival time are calculated, and the corresponding arrival event is inserted into the queue (*User Arrives at Station (to rent)*). In the example, it takes the user ten minutes (600 sec.) to reach station 55.

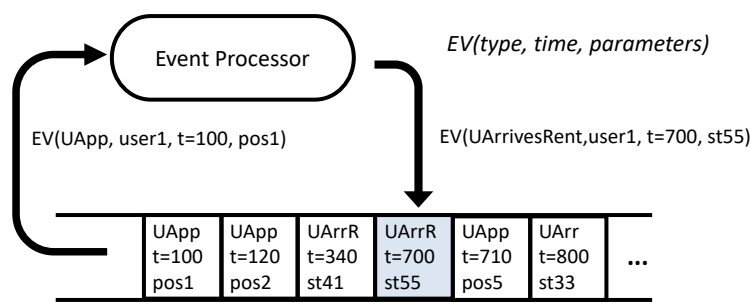

Fig. 3. Event processing example.

---

In case a user decides to reserve a bike, the event processor calculates if the established maximum reservation time (a system configuration parameter) would expire before the user arrives at the corresponding station. If that was the case, the *Bike Reservation Timeout* event would be inserted into the queue instead of a *Users Takes Bike* event.

### *4.2    User Management*

Users are the principle simulated entities. In fact, a simulation consists of simulating a set of users' behaviours in the system, since the moment they appear until they leave. As mentioned before, the simulation engine controls the flow of user events that may occur by processing the event queue. Most events require some decisions that have to be taken by the corresponding user (e.g., select a station, decide to reserve, …). In real life different users have different behaviour models and react differently under similar circumstances. For example, some people may not mind walking 500 meters to get a bike while others may only be willing to walk less than 300m. We allow this by using the notion of user types, which represent implementations of different user models.

Each *user type* must provide an implementation of the following user decision methods that the simulation engine may invoke, and which correspond to the different decisions presented in Fig. 1 but contextualised to the situation that has led to the need for a user decision (e.g., a failed bike rental, etc.):

- *decideAfterAppearning*

- *decideAfterFailedRental*

- *decideAfterFailedBikeReservation*

- *decideAfterBikeReservationTimeout*

- *decideAfterGettingBike*

- *decideAfterFailedReturn*

- *decideAfterFinishingRide*

- *decideAfterFailedSlotReservation*

In practice, we define an abstract type *User* which specifies the abstract decision methods and also provides a basic set of user parameters and default values:

- *position*: indicates the geographical position where a user appears

- *destinationPlace*: the final destination the user wants to go to

- *timeInstant*: the instant (second from simulation start) where the user appears

- *walkingVelocity*: the walking velocity of the user (default 1.4 m/s (Mohler, Thompson, Creem-Regehr, Pick & Warren, 2007; Levine & Norenzayan, 1999))

- *cyclingVelocity*: the cycling velocity of the user (default 6.0; m/s)

Besides these parameters, different user types can specify additional parameters, like the maximum number of rentals (or reservation) a user would try until he abandons the system, or the maximum distance he is willing to walk to find a bike.

An interesting issue is how users choose stations to rent or to return a bike. In particular, a user type may use a specific information or recommendation system for this task. This allows us to test and evaluate different recommendation strategies, e.g., for obtaining a better balancing of the bikes in the global system.

Currently, we have implemented the following user types:

- *Uninformed*. This user tries to take/return bikes just from the nearest station. He does not have information about availability of bikes/slots, so it may happen that there are no bikes/slots available when the user arrives at the station.

- *Informed*. It is an informed user that knows the state of the fleet and always selects the closest station with bikes or the closest station to its destination point that has available slots, when returning a bike. It may happen that the user can not rent a bike when arriving at a station, because other users have taken all available bikes before.

- *Obedient*. It contacts a particular recommender system and always follows its suggestions. This user is adequate for testing different recommendation strategies.

- *Uninformed-R*. It is like *Uninformed* but making reservations before going to stations.

- *Informed-R*. It is like *Informed* but making reservations before going to stations.

- *Obedient-R*. It is like *Obedient* but making reservations before going to stations.

  All these user types have additional parameters:

- *minRentalAttempts*: specifies the number of rental attempts the user would try before abandon the system without renting a bike,

- *maxDistanceToRentBike*: specifies the maximum distance a user is willing to walk to rent a bike. For instance, an *Informed* user would abandon if there is no station with an available bike within this distance.

- *intermediatePosition*: specifies a geographical location in the region of interest. If this parameter is not null, the user may decide to "go for a ride" after getting a bike. In this case, the user would go with the bike to the specified position. After arriving there, he would decide to return the bike.

The implementation of new user types in the simulator is fairly easy. Any new user simply has to extend the abstract *User* class. Furthermore, we use reflection to specify the user types of the users of a particular simulation. Thus, no additional changes in the simulator are necessary.

### 4.3    *Experiment Configuration*

Simulation experiments are configured by providing three types of configuration parameters: Stations, Users and Global, as shown in Fig. 4. These configuration parameters are stored in *json* files.

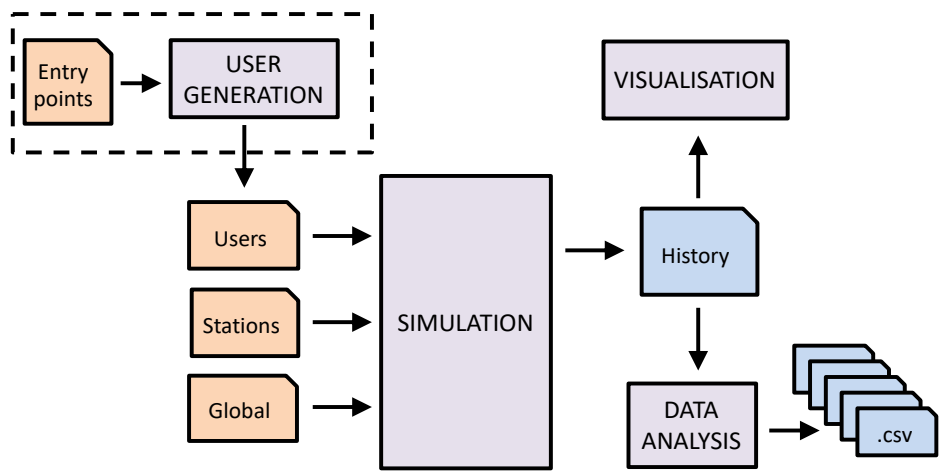

Fig. 4. Experiment configuration and generated files

*Stations.*

It contains information about the stations in the system, including physical location, capacity and initial number of bikes.

*Global.*

Specifies the global parameters of a simulation experiment. They include the following

fields:

- *reservation time*: the time before a reservation expires

- *simulation time*: the total duration of the simulation (in seconds)

- *random seed*: (optional) it can be used to generate the same sequence of random values.

- *bounding box*: of the underlying area of interest (geographical coordinates of top-left and bottom-right corners)

- *map*: specifies the file containing the map for calculating directions. We use currently OSM to calculate routes and times

- *output path*: the path where the history files are written.

- *recommendationSystemType*: (optional) is the type of recommendation system used by users that make use of such a system. It contains a *typeName* and a set of *parameters*.

With regard to the *recommendationSystemType*, similar to user types, different recommendation systems can be implemented. Each recommendation system has to extend the abstract class *RecommendationSystem* that implements some auxiliary methods and defines the abstract methods: *recommendStationToRentBike* and *recommendStationToReturnBike*. Each implementation of a recommendation system has to implement these two methods that should return an ordered list of stations which are recommended to a specific user (at a given time and position) for renting/returning a bike. Again, we use reflection to select the particular implementation of a recommendation system in the simulator, based on the value of the field *typeName* in the configuration file. Parameters of a specific recommendation system are specified in the *parameters* set.

*Users.*

Specifies the list of all users that are generated during the simulation. Each user is specified through the following parameters:

- *userType*: the user type of this user. It corresponds to an implementation of a particular user model.

- *position, destinationPlace*, *timeInstant*: the parameters all users must specify (as described previously)

- the optional parameters: *walkingVelocity* and *cyclingVelocity* as specified before

- any other parameters that are specific to the particular *userType* implementation

All three configuration files can be generated manually or via a user interface (see Fig. 5). With respect to the Users configuration, we also implemented a *User Generator* tool for generation users randomly by specifying a set of *Entry Points*. Each entry point represents a "user generator", that allows to generate random values for the fields: *position, destinationPlace*, *timeInstant*. The *timeInstant* is generated from a starting time with a Poisson distribution with a parameter λ. The Poisson distribution is a discrete probability distribution that represents events occurring at time intervals and independent of the time of the last event. It has been used by others to model user arrivals at a bike sharing system, e.g. (Chemla et al. 2013). The values for the *position* and *destinationPlace* fields are generated randomly in a circle centred at a given location and with a specified radius. The user generator creates a *User* configuration file, which can then be used as input to the simulator.

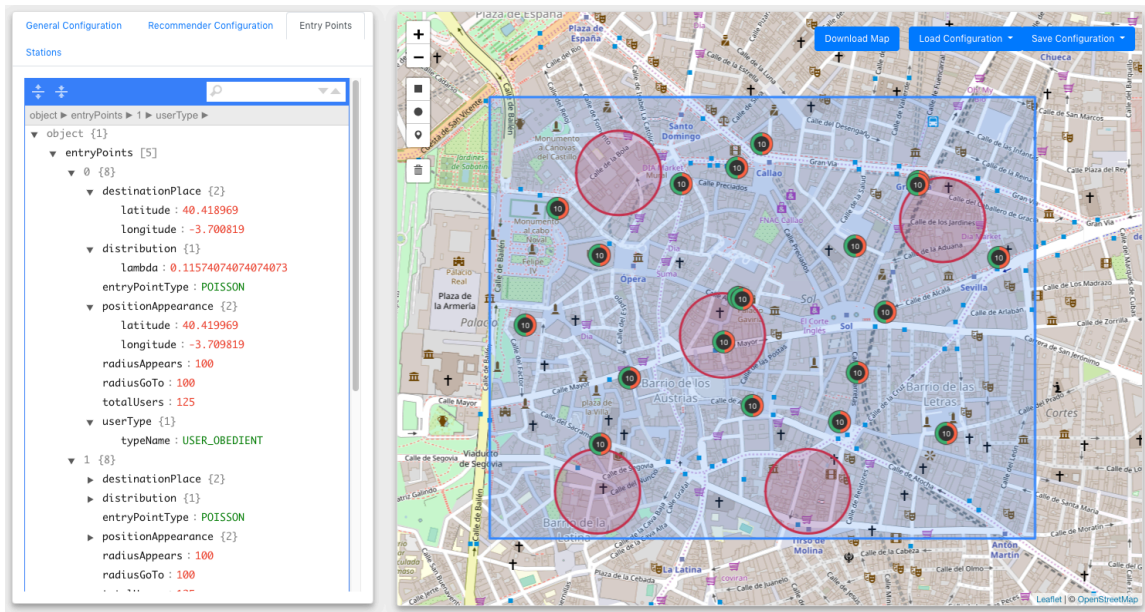

Fig. 5. Snapshot of the configuration interface. Big red circles represent entry points, with their location in the centre. Small circles are stations, with the number of available bikes in them. The ratio of available bikes and free slots is shown in red and green, respectively. The left-hand side shows the details of one entry point.

### *4.4    Visualisation*

The aim of the visualisation module is to display the produced history data of the simulation in a graphical and appealing way (see a snapshot in Fig. 6). This includes rendering geographic data on a map and presenting the current state of the system at each moment.

The visualization essentially acts as a reproducer of a set of recorded data where the entities are displayed on a map with additional status information. In particular, the situations of all stations and all users that are currently in the system is presented. The reproduction is done in a playback mode simulating real-time. Internally, the visualization module processes the data of the events that have taken place (as stored in the history files) and updates the situation of the entities in simulated real-time. In contrast to the simulation engine, here, the movements of the users (either when walking or when cycling) are simulated, that is the position of the users is continuously updated. The speed in which real-time is simulated can be controlled and a negative factor will re-wind the playback.

To render the map and to simulate movements, data from OpenStreetMap is used. The actual entities (users and stations) are displayed by interactive markers that provide additional information of the current state of each entity.

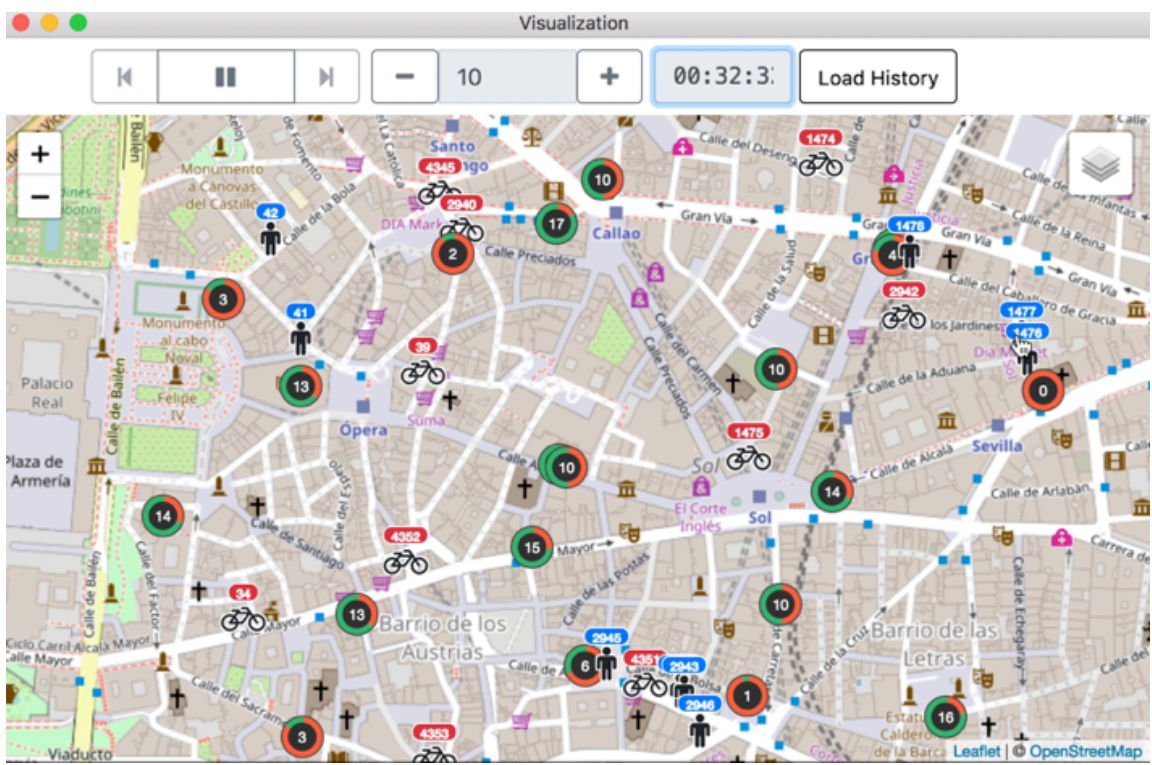

Fig. 6. Snapshot of the visualization interface. Bike and person symbols represent users riding or walking, respectively. Numbers on top of the symbols are identifiers.

### 4.5    *Data analysis*

In this section we describe the quality metrics our simulator generates to evaluate the performance of a BSS. More extended explanations can be found in [authors].

We use the following notations to define the metrics: $N$ is the total number of users, (i.e. total bike demand), *Successful hires* (*SH*) is the total number of bike rentals, *Failed* *hires* (*FH*) represents the total number of attempts to hire a bike that failed due to unavailability, *Successful returns* (*SR*) and *Failed returns* (*FR*) represent the same numbers for bike returns. Based on these measures, we calculate the following metrics:

- *Demand satisfaction* (*DS*): ratio of users who were able to hire a bike (either at first trial or not), including those who booked a bike in advance. This is one of the most common metric used to evaluates BSS (e.g. (Chemla et al., 2013))

$$DS = SH / N$$

    With regard to bike returns, we do not define a similar measure since we assume that all users that have taken a bike must return it to some station.

- *Hire efficiency* (*HE*): ratio between the number of rentals and the total rental attempts of those users who hired a bike ($FH_h$).

$$HE = SH / (SH+FH_h)$$

- *Return efficiency* (*RE*): ratio between the number of returns and the total return attempts:

$$RE = SR / (SH+FR)$$

- *Average Empty Time (AET)*: is the average time a station is empty, which means it is potentially denying a service. This metric is useful when users have access to information about bike availability before making their decision (in real life, e.g. via an app). In those cases, they may go to the nearest station with available bikes.

- *Average Deviation (AD) with regards to a balanced situation*. We consider a station is balanced if the number of available bikes is half its capacity (thus keeping half empty slots). For each station it is calculated the average deviation as the average absolute error. This metric is the average of all stations.

- *Users' time in the system*. It is the time users spend in the system. The total time ($TT$) is the sum of walking time to origin station ($T_{os}$), cycling travel time to return station ($T_{rs}$) and walking time to final user destination ($T_{fd}$):

$$TT = T_{os} + T_{rs} + T_{fd}$$

Note that $T_{os}$ and $T_{rs}$ include walking or cycling (respectively) from station to station if no available bike/slot is found.

All these measures are generated by the data analyser and are written to files in *csv* format so they can be imported into more powerful data analysis tools for further analysis.

## 5    Simulation examples

In the sequel we present three examples of the usage of the Bike3S simulator: a simulation of "designed scenarios" in the city of Madrid and simulations with real data in the cities of Madrid and London. The aim of the examples is to show the possibilities of our simulation tool for evaluating giving management decisions or strategies, especially balancing strategies. However, in this paper, our research is not focused on adequate balancing strategies. We just implemented a set of rather simple strategies as examples.

### 5.1    *Designed scenarios*

With this set of experiments, we show the possibility to create and evaluate artificial scenarios. We chose a 3km square map of central Madrid and set 20 stations in real lo-cations of the current bike-sharing system of Madrid (BiciMAD). They all have a ca-
pacity of 20 bikes. Initially, each station was set to 10 available bikes (thus, 10 empty
slots too). We set five entry points. The locations of those elements are shown in Fig. 5.
The appearance radius is 200m and destination is randomly chosen in the whole area.
We set a maximum of two failed attempts to rent or reserve a bike before leaving the
system without using it and a maximum distance of 600 meters a user would be willing
to walk in order to find a bike. Walking and cycling velocity was set to 1.4 and 6 meters
per second, respectively.

With this set of parameters, we analyse the performance of the system with dif-
ferent user types for increasing demand data (increasing number of users). We carried
out experiments where the generation ratio of users at each entry point was 10, 20, 40,
60, 80, 120 and 150 users per hour. We evaluated the following types of users (pre-
sented in section 4.2): *Uninformed, Informed, Uninformed-R, Informed-R* and three
types of *Obedient* users: i) *Obedient-AvR*, where users use a recommendation system
that recommends the station with the most available resources (bikes or slots, respec-
tively), ii) *Obedient-AvR/Dist* where the station with the best ratio between the available
resources at that station and the distance of the user to the station is recommended, and
iii) *Obedient-AvR-R,* like *Obedient-AvR* but users always reserve bikes or slots at the
recommended stations.

Fig. 7 to Fig. 12 show the evolution of the performance of the different user
types with increasing demand and for the different metrics defined in section 4.5. As ex-
pected, the performance with *Uninformed* users is the worst of all cases. *Obedient* users
outperform *Informed* users in the efficiency measures since the use of a balancing strat-
egy reduces the number of times that stations get empty. However, *Obedient-AvR* de-
creases significantly in demand satisfaction and hire efficiency with higher demands.

This is because many users that appear at almost the same moment and in similar regions will be sent to the same best station and will find that there is no bike left when they arrive (only the firsts will get them). Return efficiency is very high in almost all cases. This is due to the fact that users' destinations are not concentrated in specific locations as with appearances as well as that some users abandoned the system without hiring. Furthermore, *Obedient-AvR* present worse total time values which is due to the fact that users are send to stations that are rather far away. Both problems are mitigated with the *Obedient-AvR/Dist* type. With regard to reservations, it can be observed that users that make reservations have generally a higher chance to get a bike.

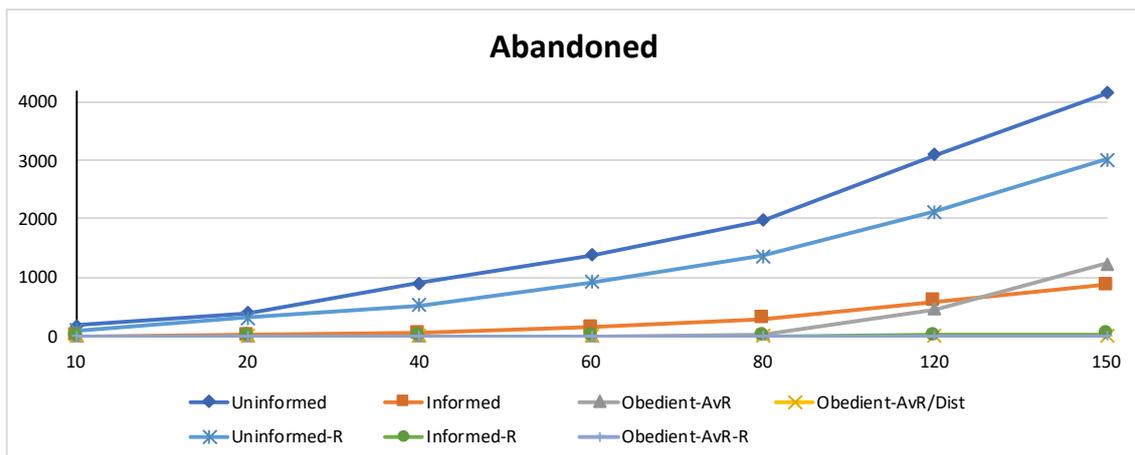

Fig. 7. Abandoned execution results. This metric is equivalent to Demand Satisfaction but is easier to see. Informed-R coincides with Obedient-AvR-R.

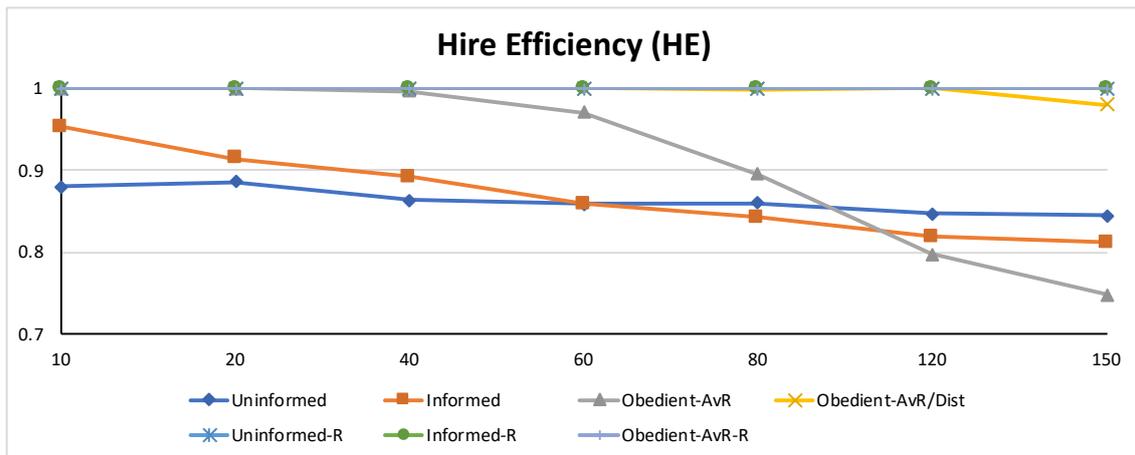

Fig. 8. Hire Efficiency execution results. Unformed-R, Informed-R and Obedient-AvR-R coincide.

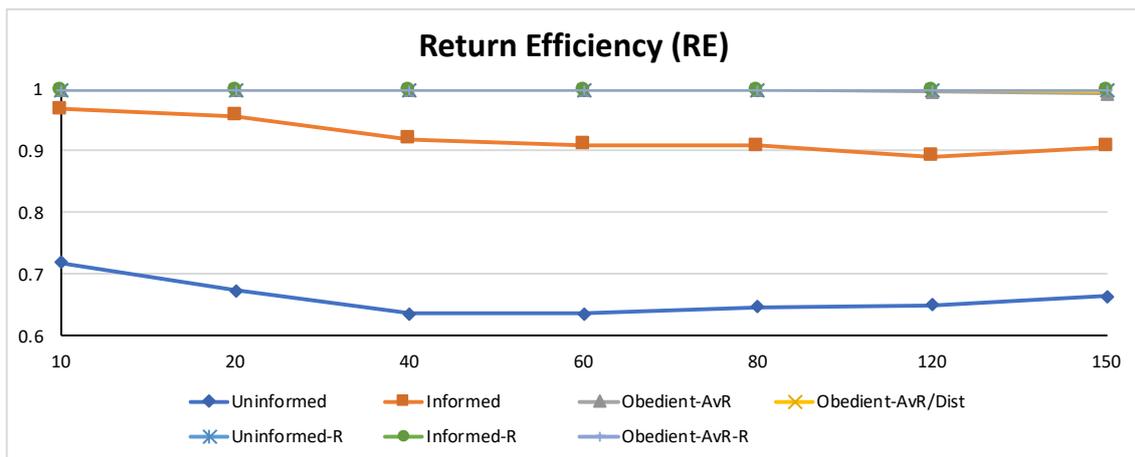

Fig. 9. Return Efficiency execution results. Unformed-R, Informed-R and Obedient-AvR-R coincide.

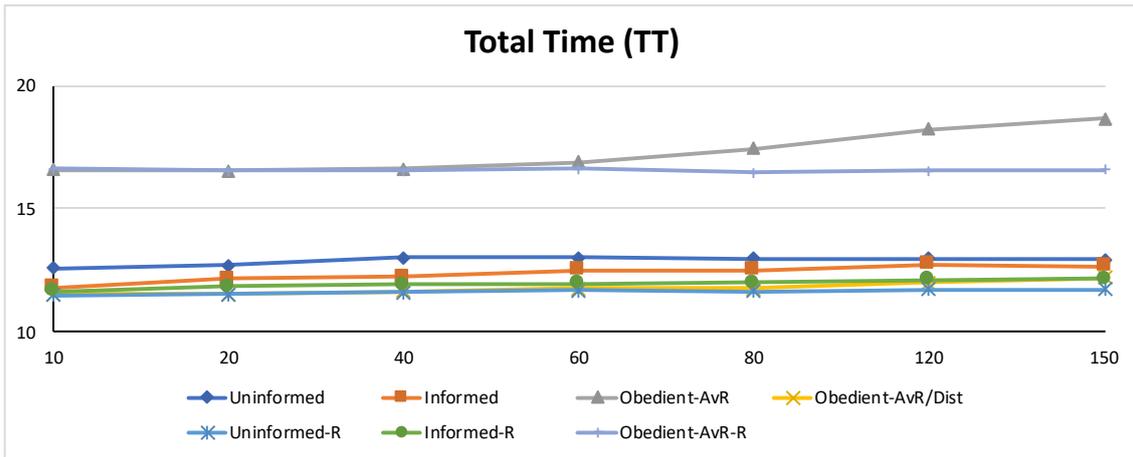

Fig. 10. Total Time execution results.

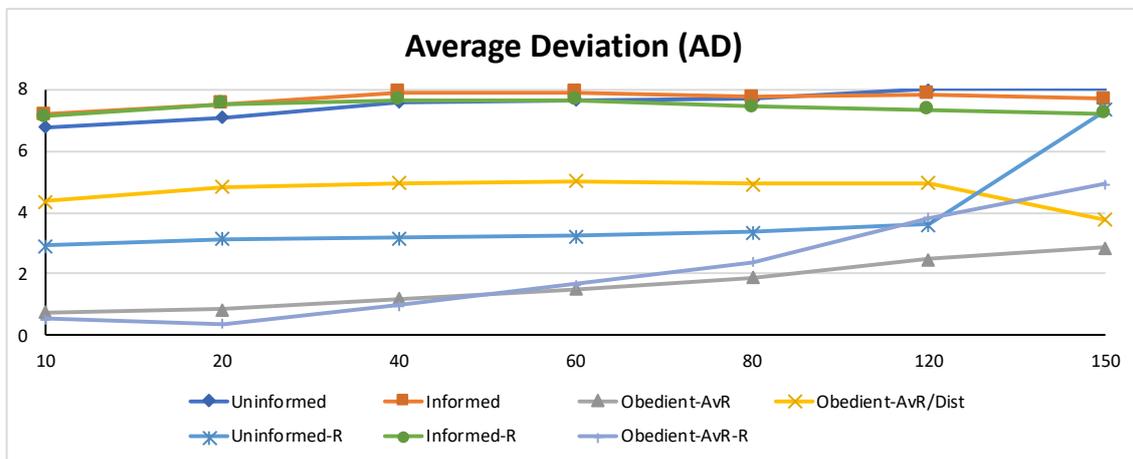

Fig. 11. Average Deviation execution results.

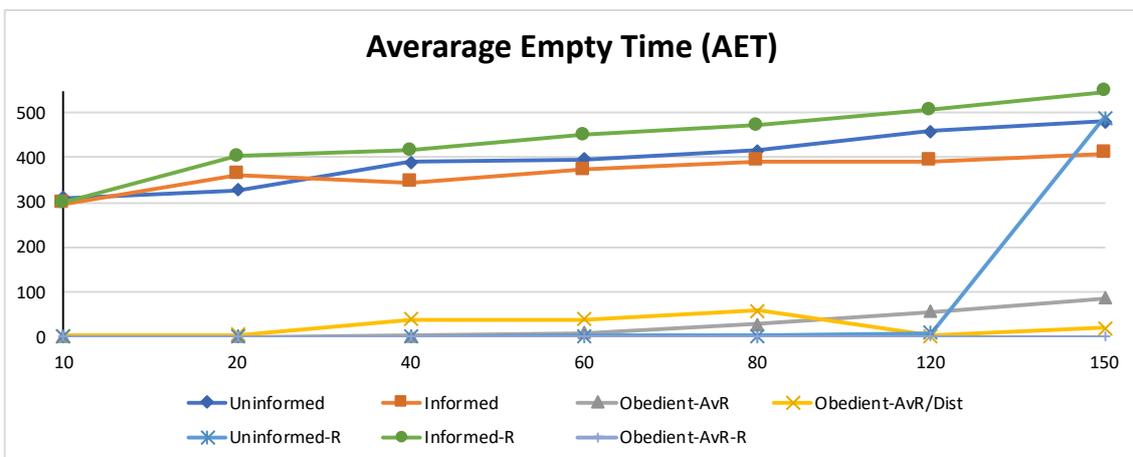

Fig. 12. Average Empty Time execution results.

## 5.2 *Real scenario simulation. Madrid and London.*

In this set of experiments, we used real data from Madrid and London. The goal is to test our simulation with different real infrastructures (number, location and size of stations) and use patterns.

In the first experiments, used data from BiciMAD, the public bike sharing system of Madrid (Spain), which covers an area of about 5x5 km square of central Madrid. BiciMAD counts on 173 stations and 1702 bikes. Each station has around 20 slots to plug in bikes.

We simulated the operation of one day of BiciMAD with real data of bike trips[3]. In particular we used the data of the 25th of June 2018 a day with a fairly high usage of the system. For each trip (13104 in total), data include time of taking a bike, origin station, time of returning, destination station and the approximate route. Origin and returning times are given just in hours (without minutes). In order to simulate the trips in a more realistic way we randomly generated appearance time in minutes within the specified hour (with a uniform distribution). In addition, we also randomly generated the appearance and destination location of users with a uniform distribution within a radius of 200 meters from the real origin and destination station, respectively. All the stations were initialised with a ratio bikes/empty slots of 0.5 (stations initial state information was not available). We set a maximum of three failed attempts to rent or reserve a bike before a user would abandon the system without using it. Furthermore, we specified a maximum radius of 600 meters as the distance a user is willing to walk in order to find a bike. If he does not find a bike at this distance, he will also abandon the system. The

---

[3] The data are publicly available at https://opendata.emtmadrid.es/Servicios-web/BICIMAD.

walking velocity of all users has been set to 1.4 meters per second and the cycling ve-
locity has been obtained from the information of the real route information.

We analysed the performance of the following user types (as explained in section 4.2):
*Uninformed*, *Informed*, *Obedient/AvR*. and *Obedient-AvR/Dist*.

Table 2 shows the results of the experiments. The efficiency data (*DS*, *HE* and
*RE*) are very high in general, especially for users that decide based on recommenda-
tions. This is normal because the historic data used in the simulation only contain suc-
cessful hires. Despite this fact, those values are not always 1 since we had to randomly
distribute users within each hour and close location as explained before, so the data are
not exactly the same as the historic records. Uninformed users get the lowest perfor-
mance in demand satisfaction (*DS*) and efficiency (*HE* and *RE*). That is due to the fact
that users go to their nearest station without even checking whether there are bikes
available. The other user behaviours present a better performance. If users follow the
recommendation strategy to go always to the station with the most resources, the results
present the best values in average empty time and deviation. The reason is that these
metrics analyse station status and this strategy tends to keep individual stations as bal-
anced as possible. However, the performance from the users' perspective, i.e. the time
users spend in the system, is much worse. If the recommendation is done by taking into
account both, the distance a user would have to walk to a station as well as the available
resources, then a reasonable balance of bikes at station is obtained, and users have a
fairly low total time in the system.

Table 2. Experiments results for Madrid. Bold numbers indicate the best obtained result for each metric.

| UserType | # abandoned | DS | HE | RE | TT (min) | AD | AET (min) |
|---|---|---|---|---|---|---|---|
| Uninformed | 200 | 0.98 | 0.93 | 0.80 | 14.6 | 5.3 | 83.2 |
| Informed | 83 | 0.99 | 0.98 | 0.94 | 14.0 | 5.4 | 91.5 |
| Obedient-AvR | **0** | **1.00** | **1.00** | **1.00** | 21.7 | **1.9** | **0.4** |
| Obedient-AvR/Dist | 18 | 1.00 | 1.00 | 0.99 | **13.8** | 4.3 | 28.6 |

The experiments using real data from London cover an area of about 20x10 km square. The system includes 783 stations and 20178 bikes. The capacity of each station varies from 10 to 62 slots. We used the data of the 30th of May 2018, which included 31736 trips. The rest of parameters and decisions are the same as in the previous experiment. The results obtained (Table 3) are in line with the ones from Madrid.

Table 3. Experiments results for London. Bold numbers indicate the best obtained result for each metric.

| UserType | # abandoned | DS | HE | RE | TT (min) | AD | AET (min) |
|---|---|---|---|---|---|---|---|
| Uninformed | 488 | 0.98 | 0.98 | 0.83 | 22.4 | 6.3 | 12.0 |
| Informed | 60 | 1.00 | 0.98 | 0.93 | **22.2** | 6.6 | 16.3 |
| Obedient-AvR | **0** | **1.00** | **1.00** | **1.00** | 30.4 | **3.6** | **0.0** |
| Obedient-AvR/Dist | **0** | **1.00** | 1.00 | 0.99 | 22.4 | 5.7 | 1.2 |

## 6    Conclusion

In this paper we have described Bike3S, a station-based bike sharing system simulator. The objective of Bike3S is to analyse the behaviour of a BSS given an infrastructure (stations and bikes) and expected demand in different points of the area of interest. In addition, the simulator can be used to evaluate balancing strategies. Bike3S is highly customisable to account for different user behaviours, infrastructure and experiment configurations. We presented several use cases to show the type of experiments that can be carried out and the results that can be obtained.

The modular design of Bike3S allowed us to separate the configuration and user generation from the simulation execution, and the latter from the visualisation and analysis. Thus, the simulator can generate users or load them from a file. Likewise, the visualisation interface or data analysis tool can load previously stored simulation histories. One of the main characteristics of Bike3S is its capability to be easily extended with new user types and balancing strategies.

We are currently working on designing incentive-based balancing strategies, which are evaluated using Bike3S. Some preliminary works can be found in [authors]. In the future, we plan to extend the simulator in several lines. One is including fares to hire bikes into the infrastructure, so that discount-based incentives or dynamic pricing balancing strategies can be implemented. Another extension includes considering electric bicycles, which implies managing battery load levels, discharge/load times, decisions on which bikes to hire, etc.

**References**

Caggiani, L., & Ottomanelli, M. (2012). A modular soft computing based method for vehicles repositioning in bike-sharing systems, *Procedia - Social and Behavioral Sciences, 54*, 675-684.

Chemla, D., Meunier, F., Pradeau, T., Calvo, R. W., Yahiaoui, H. (2013). Self-service bike sharing systems: Simulation, repositioning, pricing. https://hal.archives-ouvertes.fr/hal-00824078.

Contardo, C., Morency, C. & Rousseau, L. M. (2012). Balancing a dynamic public bike-sharing system. *Technical Report CIRRELT*, vol. 4.

Erdoğan, G., Battarra, M., Calvo, R.W. (2015). An exact algorithm for the static re-balancing problem arising in bicycle sharing systems. *European Journal of Operational Research* 245(3), 667–679.

Fellendorf, M. (1994). Vissim: A microscopic simulation tool to evaluate actuated signal control including bus priority. In: *64th Institute of Transportation Engineers Annual Meeting* (pp. 1– 9). Springer.

Forma I. A., Raviv, T. & Tzur, M. (2015). A 3-step math heuristic for the static repositioning problem in bike-sharing systems. *Transportation Research Part B: Methodological* 71, 230–47.

Fricker, C. & Gast, N. (2012). Incentives and regulations in bike-sharing systems with stations of finite capacity. arXiv preprint arXiv:12011178.

Horni, A., Nagel, K. & Axhausen, K. (Ed.). (2016). *Multi-Agent Transport Simulation MATSim*. Ubiquity Press, London.

Krajzewicz, D., Erdmann, J., Behrisch, M., & Bieker, L. (2012). Recent development and applications of SUMO - Simulation of Urban MObility. *International Journal On Advances in Systems and Measurements*, 5(3&4), 128–138.

Levine, R. V. & Norenzayan, A. (1999). The Pace of Life in 31 Countries. *Journal of Cross-Cultural Psychology*. 30 (2), 178–205.

Mohler, B. J., Thompson, W. B., Creem-Regehr, S. H., Pick, H. L. Jr. & Warren W. H. Jr. (2007). Visual flow influences gait transition speed and preferred walking speed. *Experimental Brain Research*, 181 (2), 221–228.

O'Mahony, E. & Shmoys, D. B. (2015). Data analysis and optimization for (citi)bike sharing. In *Proceedings of the Twenty-Ninth AAAI Conference on Artificial Intelligence (AAAI'15)* (pp. 687–694). AAAI Press.

Pal, A., Zhang, Y. (2017). Free-floating bike sharing: Solving real-life large-scale static rebalancing problems. *Transportation Research Part C: Emerging Technologies* 80, 92–116.

Pfrommer, J., Warrington, J., Schildbach & G., Morari, M. (2014). Dynamic vehicle redistribution and online price incentives in shared mobility systems. *IEEE Transactions on Intelligent Transportation Systems* 15(4), 1567–1578.

Romero, J. P., Moura, J. L., Ibeas, A. & Alonso, B. (2015). A simulation tool for bicycle sharing systems in multimodal networks. *Transportation Planning and Technology*, 38(6), 646-663.

Schuijbroek, J., Hampshire, R. C. & Van Hoeve, W. J. (2017). Inventory rebalancing and vehicle routing in bike sharing systems. *European Journal of Operational Research* 257(3), 992–1004.

Shaheen, S., Guzman, S. & Zhang, H. (2010). Bikesharing in Europe, the Americas, and Asia. *Transportation Research Record: Journal of the Transportation Research Board*, 2143, 159–167.

Soriguera, F., Casado, V. & Jiménez, E. (2018). A simulation model for public bike-sharing systems. In *CIT2018. Proceedings of the XIII Conference on Transport Engineering*. Transportation Research Procedia 33. 139–146.

Waserhole, A., & Jost, V. (2016). Pricing in vehicle sharing systems: optimization in queuing networks with product forms. *EURO J. Transportation and Logistics, 5*, 293-320.